\documentclass[twocolumn,english,aps,prl,superscriptaddress]{revtex4}
\usepackage{color}
\usepackage{amsmath}
\usepackage{graphicx}
\usepackage{amssymb}

\makeatletter
\@ifundefined{textcolor}{}
{%
 \definecolor{BLACK}{gray}{0}
 \definecolor{WHITE}{gray}{1}
 \definecolor{RED}{rgb}{1,0,0}
 \definecolor{GREEN}{rgb}{0,1,0}
 \definecolor{BLUE}{rgb}{0,0,1}
 \definecolor{CYAN}{cmyk}{1,0,0,0}
 \definecolor{MAGENTA}{cmyk}{0,1,0,0}
 \definecolor{YELLOW}{cmyk}{0,0,1,0}
 }

\@ifundefined{definecolor}
 {\usepackage{color}}{}
\@ifundefined{definecolor}{\@ifundefined{definecolor}
 {\usepackage{color}}{}
}{}%\documentclass[aps,prb,preprint,showpacs]{revtex4}

\newcommand{\w}{\omega}

\newcommand{\ga}{\gamma}

\makeatother

\usepackage{babel}

\begin{document}

\title{
%Surface disorder and the fate of topological-insulator surface states
The fate of topological-insulator surface states under strong disorder
}

\author{Gerald Schubert}
\affiliation{Institut f\"ur Physik, Ernst-Moritz-Arndt-Universit\"at Greifswald, 17487 Greifswald, Germany}
\affiliation{Philips Healthcare, \"Ayritie 4, 01510 Vantaa, Finland}

\author{Holger Fehske}
\affiliation{Institut f\"ur Physik, Ernst-Moritz-Arndt-Universit\"at Greifswald, 17487 Greifswald, Germany}

\author{Lars Fritz}
\affiliation{Institut f\"ur Theoretische Physik, Universit\"at zu K\"oln,
%Z\"ulpicher Stra\ss e 77,
50937 K\"oln, Germany}

\author{Matthias Vojta}
\affiliation{Institut f\"ur Theoretische Physik, Technische Universit\"at Dresden, 01062 Dresden, Germany}

%%%%%%%%%%%%%%%%%%%%%%%%%%%%%%%%%%%%%%%%%%%%%%%%%%%%%%%%%%%%%%%%%%%%%%%

\begin{abstract}
Three-dimensional topological insulators feature Dirac-like surface states which are
topologically protected against the influence of weak quenched disorder. Here we
investigate the effect of surface disorder
% and its signatures in electronic spectra
beyond the weak-disorder limit using large-scale numerical simulations.
We find two qualitatively distinct regimes: Moderate disorder destroys the Dirac cone and
induces diffusive metallic behavior at the surface. Even more remarkably, for strong
surface disorder a Dirac cone reappears, as new weakly disordered ``surface'' states
emerge in the sample {\em beneath} the disordered surface layer, which can be understood
in terms of an interface between a topological and an Anderson insulator.
Together, this demonstrates the drastic effect of disorder on topological surface states, which
cannot be captured within effective two-dimensional models for the surface states alone.
%Our results are of immediate relevance to spectroscopic experiments on topological insulators.
\end{abstract}

\date{\today}


\maketitle
%%%%%%%%%%%%%%%%%%%%%%%%%%%%%%%%%%%%%%%%%%%%%%%%%%%%%%%%%%%%%%%%%%%%%%%

Topological insulators (TIs) are a novel form of insulators, where topological properties
of the band structure lead to the formation of metallic states at the surface
\cite{tirev1,tirev2}. Two-dimensional (2d) TIs were predicted \cite{Kane2005,
Bernevig2006} and then realized in quantum-well heterostructures \cite{Koenig2007}.
Subsequently, three-dimensional (3d) generalizations were discussed theoretically
\cite{Fu2007, Moore2007, Roy2009} and realized afterwards, primarily in Bi-based
compounds \cite{Hasan2009, Bruene2011}. 3d TIs are characterized by a set of topological
quantum numbers, which lead to a classification into strong and weak TIs.

Strong TIs possess low-energy surface states which can be described in terms of massless
2d Dirac electrons, with an odd number of Dirac cones per surface. Remarkably, these
surface states are topologically protected against the influence of disorder due to their
helical nature:
The spin--momentum locking of the Dirac theory implies that the states $\vec{k}$ and
$-\vec{k}$ have opposite spin. This in turn suppresses the lowest-order matrix element
for backscattering, $\vec{k}\rightarrow-\vec{k}$, leading to the absence of weak localization
for weak potential disorder.
What about strong disorder?
Theoretical results obtained within 2d continuum Dirac models suggest that such
states remain metallic even for arbitrarily large disorder: The beta-function
describing the renormalization-group flow of the conductivity is always
positive \cite{beenakker07,ryu07,mirlin10}.
However, by construction the applicability of these effective theories to TI surface
states is restricted to energies much smaller than the bulk gap, $\Delta$, and the effect
of surface disorder with a strength $\gamma \gtrsim \Delta$ can therefore not be captured.
Intuitively, this is because strong impurities will influence both surface and bulk
states, causing a non-trivial coupling between the two which cannot be described by a
theory for surface states alone.

To the best of our knowledge, the effect of strong surface disorder has only been discussed locally
for isolated impurities \cite{balatsky11}, but the global fate of TI surface states under
moderate or strong disorder is not known. This issue is not only of fundamental interest
in order to understand the range and limitations of ``topological protection'', but is
also of enormous practical relevance:
Impurity potentials can easily be in the energy range of 1\,eV, thereby exceeding typical
band gaps of 3d TIs, e.g. 0.3\,eV for Bi$_2$Se$_3$. Moreover, surfaces are often
intrinsically disordered, and in addition can be easily disordered intentionally to perform
controlled studies of disorder effects.
Therefore, non-perturbative theoretical studies of surface disorder are clearly called for.

\begin{figure*}[t]
\includegraphics[width=0.98\textwidth]{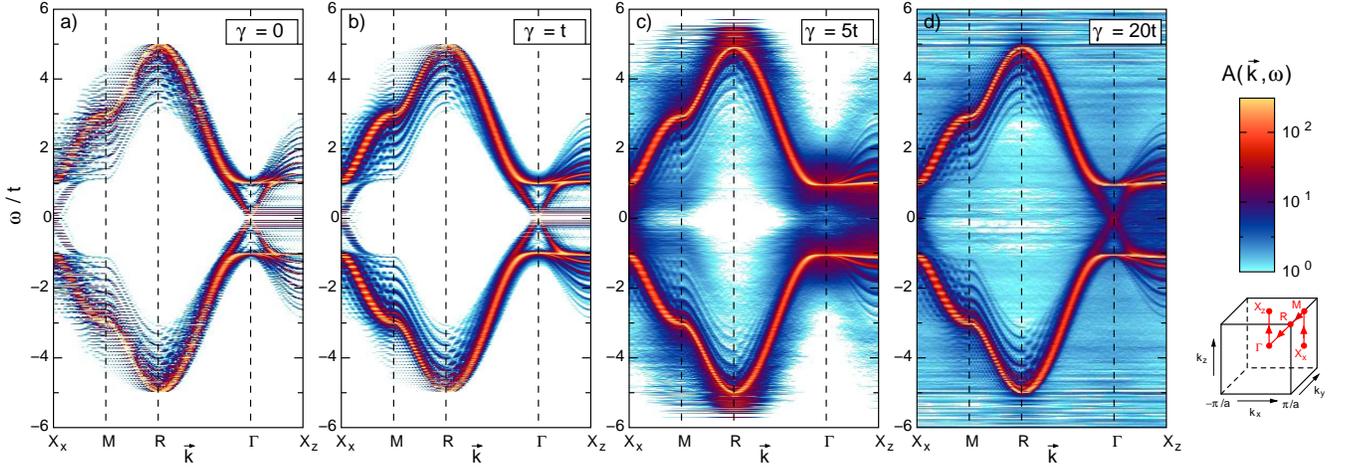}
\caption{(Color online)
Spectral function $A(\vec{k},\w)$, Eq.~\eqref{A}, along a path in the 3d Brillouin zone,
obtained by ED for the TI model \eqref{H} on a $16\times16\times33$ lattice with open boundaries.
%\hf{Data obtained by ED, histogram binning $\pi/(100a)\times 0.01t$.}
Histogram binning has been employed on the $\w$ axis, with bin width $t/50$.
The strength of the
surface disorder $\ga$ increases from a) to d). In the clean case, $\ga=0$,
bulk states exist for $|\w|\geq t$, while surface Dirac cones appear near surface momenta
$(0,0)$, as is nicely visible e.g. near $\vec{k}\!=\!\Gamma$ \cite{weightfoot}.
(The small gap at $\Gamma$ is a finite-size effect, while the ``shadow'' bands originate
from the open boundary conditions.)
The surface Dirac cones are destroyed for moderate disorder, $\ga=5t$, but re-appear for large
disorder, $\ga=20t$.
}
\label{Fig:fig1}
\end{figure*}

In this paper, we present a comprehensive study of the effect of surface disorder on 3d
TIs by means of large-scale numerical simulations of a 3d lattice model.
%
% We implement disorder on the surface but leave the bulk untouched.
Our central results can be summarized as follows:
Topological protection is only operative for disorder strengths which are small
compared to the bulk gap, $\gamma\ll\Delta$. For moderate disorder,
$\gamma\sim\Delta$, the surface
layer instead shows diffusive metallic behavior, and the momentum-space Dirac-like
structure of its states is lost. Moreover, low-energy states start to significantly
penetrate into the sample.
For strong disorder, $\gamma\gg\Delta$, states in the surface layer become strongly localized,
but new weakly disordered metallic states appear which reside primarily in the first inward
layer of the sample. Remarkably, these states display a clear Dirac dispersion,
suggesting that they can be interpreted as metallic states at the interface between a TI
and a strong Anderson insulator (AI).

Our calculations demonstrate that already moderate disorder on TI surfaces
requires to consider physics beyond effective 2d Dirac models. On the application side,
we propose that strong disorder can be used for nanostructuring of TI devices:
Disordered layers in a TI crystal act as atomic-scale barriers which cut the TI.

%%%%%%%%%%%%%%%%%%%%%%%%%%%%%%%%%%%%%%%%%%%%%%%%%%%%%%%%%%%%%%%%%%%%%%

\textit{Model and Methods.}
As a minimal model for a 3d TI we take four bands in a cubic lattice of $N$ sites,
which may also be considered as an effective model for strained 3d HgTe and for insulators of the
Bi$_2$Se$_3$ family \cite{Fu2007,Dai2008,sitte11}. The %\hf{$4N \times 4N$}
Hamiltonian can conveniently be expressed in terms of the identity, $\Gamma^0$ and four Dirac matrices
$\Gamma^a$, $\Gamma^{(1,2,3,4)} = (\openone \otimes s_z, -\sigma_y \otimes s_x, \sigma_x
\otimes s_x, -\openone \otimes s_y)$
where $s$ ($\sigma$) Pauli matrices refer to orbital (spin) space:
\begin{align}
\label{H}
\mathcal{H}
=& -t \sum_{n, j=1,2,3} \left( \Psi^\dagger_{n+\hat e_j} \frac{\Gamma^{1} - \mathrm{i} \Gamma^{j+1}}{2}  \Psi^{\phantom{\dagger}}_{n}  + \mathrm{H.c.} \right) \nonumber \\
 & +  \sum_n \Psi^\dagger_n  (\epsilon_n \Gamma^{0} +
 m\Gamma^{1})\Psi^{\phantom{\dagger}}_n\,.
\end{align}
$\Psi_n$ is a four-component spinor at site $n$, $t$ is the hopping amplitude and $m$ a parameter
which tunes the band structure.
The model describes a weak TI with two Dirac cones per surface for $|m|<t$, a strong
TI with a single Dirac cone per surface for $t<|m|<3t$, and a conventional band insulator
for $|m|>3t$. Our numerical results below are obtained for $m=2t$, where the bulk band
gap is $\Delta=2t$.

We implement quenched disorder via random on-site energies $\epsilon_n$ drawn from a
uniform box distribution with width $\ga$. For most of the paper, we will be concerned
with surface disorder, i.e., $\epsilon_n \in \left[-\ga/2,\ga/2\right]$ for
sites $n$ on each surface of the sample, and $\epsilon_n=0$ otherwise.

%%%%%%%%%%%%%%%%%%%%%%%%%%%%%%%%%%%%%%%%%%%%%%%%%%%%%%%%%%%%%%%%%%%%%%

Using state-of-the-art exact
diagonalization (ED) routines~\cite{ed_buch} we calculate
$\{ |m\rangle\}$, the two-fold Kramers degenerate eigenstates
of $\mathcal{H}$~(\ref{H}).
Those can be visualized in both momentum and position space,
via the momentum-resolved spectral function
\begin{equation}
\label{A}
  A(\vec{k},\omega) = \sum\limits_{j=1}^{4}\sum\limits_{m=1}^{4N}
  |\langle m | \psi({\vec k},j)\rangle|^2 \delta(\omega-\omega_m)
\end{equation}
and the local density of states (LDOS)
\begin{equation}
\label{rho}
  \rho_n(\omega) =  \sum\limits_{j=1}^{4}\sum\limits_{m=1}^{4N}
  |\langle m | \psi({\vec r}_n,j)\rangle|^2 \delta(\omega-\omega_m)\,.
\end{equation}
Here $\langle \vec{e}^{\,(s)}_n\!\!\otimes\!\vec{e}^{\,(b)}_j |
\psi({\vec k},j')\rangle =
\exp(\mathrm{i} \langle {\vec k}|  \vec{e}^{\,(s)}_n  \rangle) \delta_{jj'}$
is a Bloch state and
 $\langle  \vec{e}^{\,(s)}_n\!\!\otimes\vec{e}^{\,(b)}_j |
 \psi({\vec r_{n'}},j')\rangle = \delta_{nn'} \delta_{jj'}$
a Wannier state, where $|\vec{e}^{\,(s)}_n\rangle$ and $|\vec{e}^{\,(b)}_j\rangle$
denote the canonical basis vectors of position and bandindex space, respectively.

%%%%%%%%%%%%%%%%%%%%%%%%%%%%%%%%%%%%%%%%%%%%%%%%%%%%%%%%%%%%%%%%%%%%%%%

\begin{figure*}[t]
\includegraphics[width=0.78\textwidth]{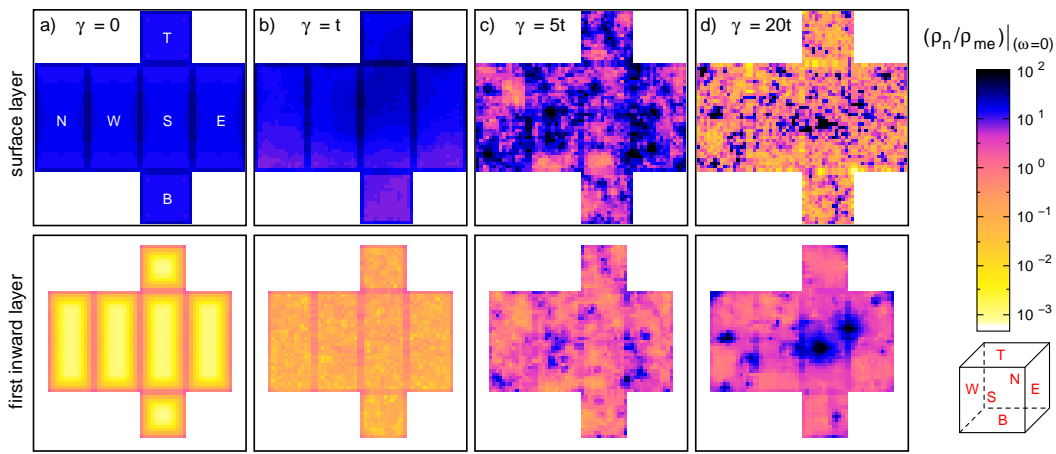}
\caption{(Color online)
Spatially resolved LDOS \eqref{rho} in the middle of the bulk gap, $\rho_n(\w\!=\!0)$,
shown for the surface (top row) and the first inward layer (bottom row) of a
$16\times16\times33$ lattice \cite{avgfoot}.
$\rho$ is normalized to the mean DOS of an extended {\em bulk} state,
$\rho_{\text{me}}=1/(4N)$, such that the weight of surface states scales as $N^{1/3}$.
Each panel shows
the 6 faces according to the scheme in the lower right corner; the strength of the
surface disorder $\ga$ increases from left to right. While in the clean case, $\ga=0$,
surfaces states penetrate the bulk only slightly, this changes already for moderate
disorder, $\gamma=5t$. For $\gamma=20t$ tendencies of localization on the outer shell are
visible, whereas weakly disordered states appear in the first inward shell.
}
\label{Fig:fig2}
\end{figure*}

\textit{Momentum-resolved spectral function.}
In Fig.~\ref{Fig:fig1}, we start by presenting results for $A(\vec{k},\w)$ along a path
following the high-symmetry directions of the {\em bulk} Brillouin zone.
In the clean case, Fig.~\ref{Fig:fig1}a, bulk states exist at energies $|\w|\geq t$,
and sub-gap surface states are clearly visible: For the model \eqref{H} with $m=2t$ the
Dirac cone is located at the surface momentum $\Gamma_s=(0,0)$ and consequently shows up
in $A(\vec{k},\w)$ at 3d momenta $\vec{k}$ which equal $\Gamma_s$ after projection onto
the respective surface, e.g.,  at $\Gamma$ \cite{weightfoot}.

While for weak surface disorder, Fig.~\ref{Fig:fig1}b, the spectral function has
changed rather little (apart from minor broadening effects), the picture is drastically altered
for moderate disorder in Fig.~\ref{Fig:fig1}c, where for $\ga=5t$ the typical
$\epsilon_n$ is somewhat larger than the bulk gap. As expected, the bulk states are well
defined, but any sub-gap states have lost their momentum-space structure. As will become
clear below, the surface states have acquired diffusive character.
Finally, Fig.~\ref{Fig:fig1}d is astonishing, as it shows that sub-gap states with a
well-defined Dirac dispersion re-appear for strong surface disorder, $\ga=20t$.

%%%%%%%%%%%%%%%%%%%%%%%%%%%%%%%%%%%%%%%%%%%%%%%%%%%%%%%%%%%%%%%%%%%%%%%

\begin{figure}[b]
\includegraphics[width=0.47\textwidth,clip]{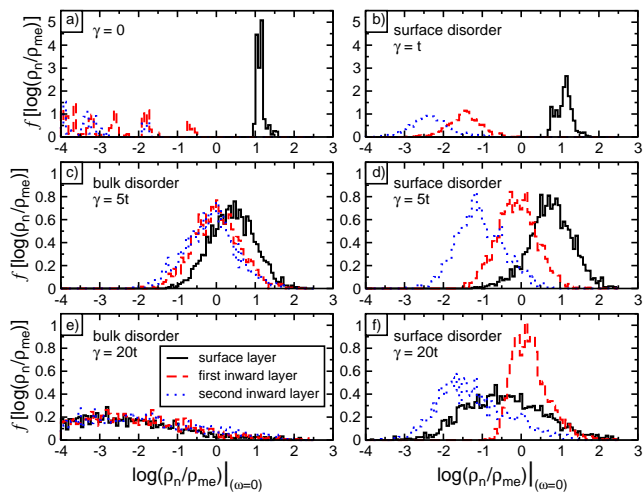}
\caption{(Color online)
Distribution of the in-gap LDOS, $\rho_n(\w\!=\!0)/\rho_{\text{me}}$, for the
surface and two next inward layers. Panels b,d,f) show different strengths
$\ga$ of surface disorder, and panels c,e) show bulk disorder, i.e., all $\epsilon_n \in
\left[-\ga/2,\ga/2\right]$ for comparison.
While the clean case a) has sizeable low-energy LDOS only on the outer layer, weight is
distributed to the inner layers with increasing disorder. For strong surface disorder,
panel f), the broad distribution for the outer layer indicates localization, while
disorder effects are much weaker on the high-intensity first inward layer.
In contrast,
for bulk disorder the broad LDOS distribution essentially coincides for all layers
(surface states are destroyed) and indicates diffusive metallic behavior in c) and
localization in f).
}
\label{Fig:fig3}
\end{figure}

%%%%%%%%%%%%%%%%%%%%%%%%%%%%%%%%%%%%%%%%%%%%%%%%%%%%%%%%%%%%%%%%%%%%%%%

\begin{figure*}
\includegraphics[width=0.99\textwidth]{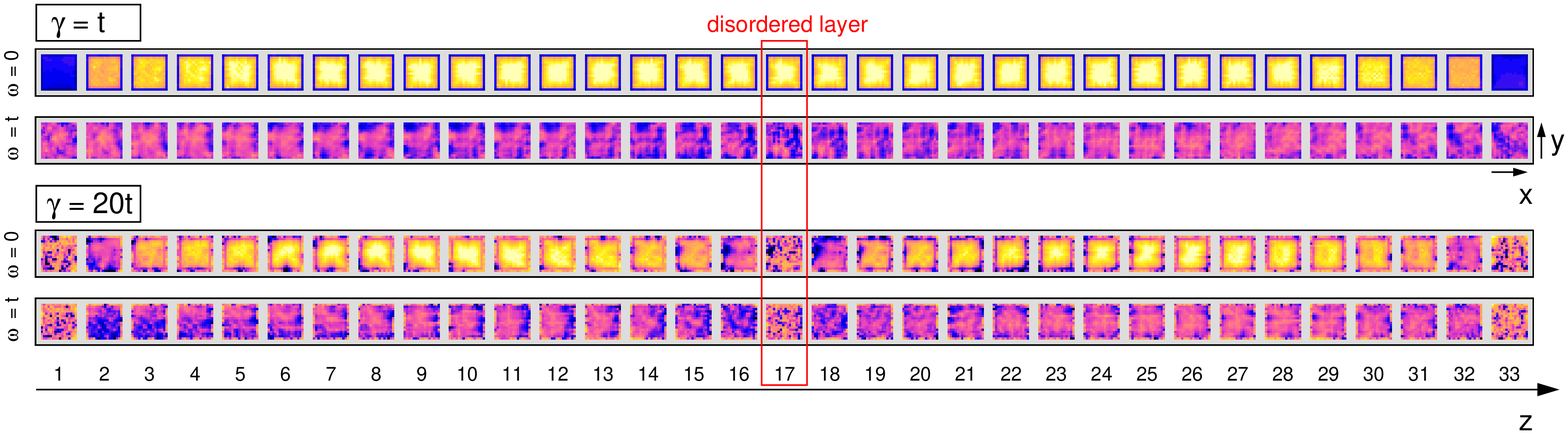}
\caption{(Color online)
Spatially resolved LDOS at energies $\w=0$ and $t$ for a $16\times16\times33$ sample with
both surface and midplane disorder, for disorder strengths $\gamma=t$ (top) and
$\gamma=20t$ (bottom). The individual panels show the LDOS slice-by-slice for the entire
sample. While weak disorder leaves the physics of the clean sample untouched, strong
disorder leads to localization on the surfaces and in the midplane, and new weakly
disordered sub-gap states appear both in layers 2 and 32 {\em and} in layers 16 and 18
adjacent to the midplane: The disordered midplane has cut the TI into two.
%
%\todo{Add schematic 3d pic which shows the ``divided'' sample?}
}
\label{Fig:fig4}
\end{figure*}

%%%%%%%%%%%%%%%%%%%%%%%%%%%%%%%%%%%%%%%%%%%%%%%%%%%%%%%%%%%%%%%%%%%%%%%

\textit{Local density of states.}
More information can be extracted from the LDOS $\rho_n(\w)$ \cite{avgfoot}.
Fig.~\ref{Fig:fig2} displays the LDOS for the two outer layers (surface and first inward
layer) of a finite system at $\w=0$, i.e., in the middle of the bulk gap \cite{ldosfoot}.
$\rho_n(\w)$ can further be used to construct the LDOS probability distribution $f[\rho_n(\w)]$ \cite{ssbfv10}, i.e., a
histogram of LDOS values at a fixed energy $\w$, for specific parts of the sample.
Fig.~\ref{Fig:fig3} shows such layer-resolved histograms for $\rho_n(\w\!=\!0)$.

In the clean case, Figs.~\ref{Fig:fig2}a and \ref{Fig:fig3}a, we have homogeneous surface
states, with the weight being located almost exclusively in the outermost layer. These
states are only weakly perturbed for $\ga=t$, Figs.~\ref{Fig:fig2}b and \ref{Fig:fig3}b,
consistent with topological protection (although the weight on the
first inward layer is seen to increase).

Moderate surface disorder, $\ga=5t$, significantly re-distributes part of the weight of the
in-gap states from the surface to the first inward layer (and even to
the second inward layer),
Figs.~\ref{Fig:fig2}c and \ref{Fig:fig3}d. Moreover, the states become inhomogeneous, and
the LDOS distribution significantly broadens. These are clear signatures of diffusive
metallic behavior -- this appears consistent with the fact that a 3d TI under influence
of moderate {\em bulk} disorder becomes a diffusive metal \cite{nomura11,guo10}.
(The latter is nicely visible in Fig.~\ref{Fig:fig3}c, which shows disordered low-energy
states in all layers for a bulk $\ga=5t$.)
We conclude that, for surface disorder, low-energy surface states do exist for $\ga=5t$,
but their momentum-space Dirac structure is lost, consistent with % the result in
Fig.~\ref{Fig:fig1}c.

We now turn to strong surface disorder, $\ga=20t$. Here, the main weight of the
low-energy states has been pushed into the sample (!) to the first inward layer,
Fig.~\ref{Fig:fig3}f. In contrast, the LDOS in the surface layer is small with a very
broad distribution, corresponding to the ``fragmented'' real-space structure in
Fig.~\ref{Fig:fig2}d. The surface layer thus shows a clear tendency towards strong
Anderson localization. (This conclusion is again supported by the corresponding
result for bulk disorder, Fig.~\ref{Fig:fig3}e, signalling bulk AI behavior \cite{locfoot}.)
At the same time, the first inward layer displays a much narrower LDOS distribution, i.e., is
only weakly affected by disorder (recall that the $\epsilon_n$ are zero here). Together
with the data in Fig.~\ref{Fig:fig1}d, this shows that metallic low-energy states have
emerged in the first inward layer and exhibit a well-defined Dirac structure in momentum
space, much like the surface states of the clean sample.
The situation thus can be interpreted as an interface between a TI and an AI (which here
consists of the surface layer only), with TI interface states forming in the
outermost layer of the clean TI.

In the limit of large $\gamma$, the behavior can be rationalized by noting that the
surface sites with $|\epsilon_n|\gg t$ effectively decouple from the bulk, their only
effect being to introduce an effective disorder in the first inward layer of strength
$t^2/\gamma$, i.e., a smaller TI with weak surface disorder obtains.

%%%%%%%%%%%%%%%%%%%%%%%%%%%%%%%%%%%%%%%%%%%%%%%%%%%%%%%%%%%%%%%%%%%%%%%

{\it Relation to surface Dirac theories.}
Given the drastic effects of disorder on TI surface states, a comment on disorder effects
in 2d Dirac theories is in order. Recent work \cite{beenakker07,ryu07,mirlin10} has shown
that even strong disorder {\em cannot} localize electrons of a single Dirac cone, which
has lead to the notion of a supermetal and self-organized criticality in the presence of
interactions. However, these theories are constructed in the continuum such that the
cutoff energy of the Dirac spectrum is the largest energy scale. Therefore lattice
effects are not captured, and it is precisely lattice effects which are responsible for
strong localization in the limit of strong on-site disorder. (For TIs, such lattice
effects include the coupling to sites away from the surface layer.)

We conclude that the applicability of 2d Dirac theories to TI surface states is limited
to energy scales small compared to the bulk gap $\Delta$, and thus our results are not in
contradiction to those of Refs.~\onlinecite{beenakker07,ryu07}.
Parenthetically, we note that attempts to study 2d {\em lattice} models with a single Dirac
cone are usually hampered by the fermion doubling problem which can only be circumvented
in special cases with broken time-reversal symmetry \cite{Haldane1988}.
Therefore, treating the full 3d model as done here is indispensible.

%%%%%%%%%%%%%%%%%%%%%%%%%%%%%%%%%%%%%%%%%%%%%%%%%%%%%%%%%%%%%%%%%%%%%%%

\textit{Midplane disorder: How to cut a TI.}
We can use the insights gained above for an ``experiment'': If a single strongly
disordered layer, simulating an AI, produces new Dirac interfaces states
in the TI, this can be used to effectively {\em cut} a TI.
To this end, we introduce on-site disorder selectively in one central layer (or midplane)
of the TI, Fig.~\ref{Fig:fig4}. If this disorder is strong, $\ga=20t$, the midplane shows
signatures of Anderson localization, and new low-energy states appear to the left and
right of the midplane. From Fig.~\ref{Fig:fig4} one can deduce that these layers show the
same characteristics as the layers adjacent to the disordered surfaces.

Our simulated sample thus represents a layered structure AI--TI--AI--TI--AI (the two
outer AI on the sample surface), with interface states bounding both pieces
of TI. (We have checked that the properties remain robust if the number of disordered
layers is increased.)
This suggests efficient nanostructuring via disorder: A large TI crystal can be ``cut''
into multiple TIs, preserving the single-crystalline structure, by intentionally disordering
a few atomic layers, e.g., via irradiation, which then form barriers with new interface
states.

%%%%%%%%%%%%%%%%%%%%%%%%%%%%%%%%%%%%%%%%%%%%%%%%%%%%%%%%%%%%%%%%%%%%%%%

\textit{Conclusions.}
By large-scale numerics we have demonstrated that surface disorder can strongly
disturb surface states of 3d TIs, such that their description in terms of a 2d Dirac
theory breaks down: Topological protection is only relevant for weak
disorder, while disorder with a strength comparable to or larger than the bulk gap first
leads to diffusive metallic behavior and then to Anderson localization at the surface. In
the latter case, the emerging interface between a TI and an AI comes with new low-energy
interface states, with properties similar to the surface states of a clean TI interfacing
vacuum -- the physics of these interface states clearly deserves further studies.

STM experiments have already detected strong perturbations of surface states by isolated
impurities \cite{kapi11}, and experiments on more disordered surfaces can be used to confirm our
predictions.

%%%%%%%%%%%%%%%%%%%%%%%%%%%%%%%%%%%%%%%%%%%%%%%%%%%%%%%%%%%%%%%%%%%%%%%

% \acknowledgments

We thank A. Altland, F. Evers, and C. Timm for discussions.
This research was supported by the DFG through SPP1459, SFB608, FOR960, and FR2627/3-1.

%%%%%%%%%%%%%%%%%%%%%%%%%%%%%%%%%%%%%%%%%%%%%%%%%%%%%%%%%%%%%%%%%%%%%%%

%\vspace*{-15pt}


\begin{thebibliography}{99}
\bibitem{weightfoot}
Surface states have reduced weight in $A(\vec{k},\w)$ as compared to bulk states, as
they only occupy $L^2$ sites in a sample of size $L^3$.
\bibitem{avgfoot}
Results are derived from a single realization of disorder to emphasize the typical
behavior of an individual sample.
\bibitem{locfoot}
Our samples are too small for an accurate detection of Anderson localization.
\end{thebibliography}
\end{document}